\documentclass[aps,prl,twocolumn,superscriptaddress,showpacs]{revtex4}
\usepackage[normalem]{ulem}
\usepackage{float}
\usepackage[usenames]{color}
\usepackage{graphicx}
\usepackage{ulem}
\usepackage{amsfonts}
\usepackage{amsmath}
\renewcommand{\d}{{\rm d}}
\newcommand{\lc}{\ell_{\rm c}}
\newcommand{\lp}{\ell_{\rm p}}
\newcommand{\ton}{\tau_{\rm on}}
\newcommand{\toff}{\tau_{\rm off}}
\newcommand{\ooff}{\omega_{\rm off}}

\newcommand{\teq}{\tau_{\rm eq}}
\newcommand{\muth}{\mu_{\rm th}}

\newcommand{\kT}{k_{\rm B}T}

\begin{document}

\title{Cross-link governed dynamics of biopolymer networks}

\author{Chase P. Broedersz}
\affiliation{Department of Physics and Astronomy, Vrije Universiteit, Amsterdam, The Netherlands}
\author{Martin Depken}
\affiliation{Department of Physics and Astronomy, Vrije Universiteit, Amsterdam, The Netherlands}
\author{Norman Y. Yao}
\affiliation{Department of Physics, Harvard University, Cambridge, MA 02138, U.S.A.}
\author{Martin R. Pollak}
\affiliation{Renal Division, Departments of Medicine, Brigham and Women's Hospital and Harvard Medical School, Boston, Massachusetts, U.S.A.}
\author{David A. Weitz}
\affiliation{Department of Physics, Harvard University, Cambridge, MA 02138, U.S.A.}
\affiliation{School of Engineering and Applied Sciences, Harvard University,
Cambridge, MA 02138, U.S.A.}
\author{Frederick C. MacKintosh}
\affiliation{Department of Physics and Astronomy, Vrije Universiteit, Amsterdam, The Netherlands}

\date{\today}

\begin{abstract}
Networks of stiff biopolymers cross-linked by transient linker proteins exhibit complex stress relaxation, enabling network flow at long times. We present a model for the dynamics controlled by cross-links in such networks. We show that a single microscopic timescale for cross-linker unbinding leads to a broad spectrum of macroscopic relaxation times and a shear modulus $G\sim\omega^{1/2}$ for low frequencies $\omega$. This model quantitatively describes the measured rheology of actin networks cross-linked with $\alpha$-Actinin-$4$ over more than four decades in frequency.
\end{abstract}


\maketitle
\begin{figure}
\begin{center}
\includegraphics[width=\columnwidth]{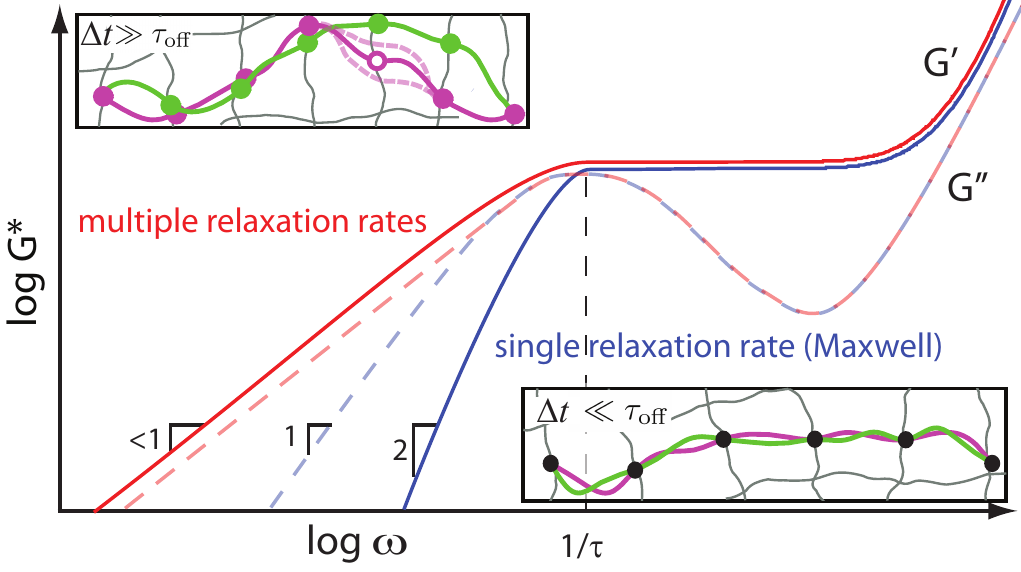}
\caption{\label{fig:model} (Color online) A schematic of the frequency dependent  shear modulus $G^*=G'+\imath G''$.  Non-permant networks can exhibit a response ranging from a single timescale ($\tau$) Maxwell-like behavior (blue lines) to a powerlaw regime with an exponent $<1$  governed by a broad distribution of relaxation times ($>\tau$) (red lines).  Upper inset: for times  longer than the unbinding time $\toff$, large scale conformational relaxation can occur via linker unbinding  (open cirlce) and subsequent rebinding at a new location. Lower inset: for shorter times, only small-scale bend fluctuations between cross-links can relax, resulting in a plateau in $G'$ for frequencies $>1/\toff$.}
\end{center}\vspace{-0.2in}
\end{figure}

Reconstituted biopolymers such as actin are excellent models for semi-flexible polymers, with network mechanics and dynamics that are strikingly different from flexible polymer networks~\cite{janmey_resemblance_1990,mackintosh_elasticity_1995, gittes_dynamic_1998, morse_viscoelasticity_1998, hinner_b.__and_tempel_m.__and_sackmann_e.__and_kroy_k.__and_frey_e._entanglement_1998, gardel_elastic_2004,koenderink_high-frequency_2006,tharmann_viscoelasticity_2007}. One essential feature setting biopolymer networks apart from rubber-like materials is the intrinsic dynamics of their cross-links. Such systems represent a distinct class of polymeric materials whose long-time dynamics are not governed by viscosity or reptation~\cite{doi_theory_1999}, but rather, by the transient nature of their cross-links. This can give rise to a complex mechanical response, particularly at long times, where the network is expected to flow. Such flow can have important implications for cells, where their internal networks are constantly remodeling, reflecting the transient nature of their cross-links~\cite{stamenovic_cell_2006}. The simplest possible description of a material that is elastic on short  timescales while flowing on long timescales is that of a Maxwell fluid; this exhibits a single relaxation time $\tau$, as depicted in Fig.~\ref{fig:model}. Indeed, some recent experiments on transient networks have suggested the existence of a single relaxation time~\cite{lieleg_transient_2008}; by contrast, other experiments---probing longer time-scales---evince a more complex viscoelastic behavior~\cite{ward_dynamic_2008, wachsstock_cross-linker_1994}, indicative of multiple relaxation times. Thus, the basic physical principles governing such transient networks remain a mystery. A predictive theoretical model is essential to elucidate the effect of dynamic cross-linking, and to help explain the complex viscoelastic behavior observed experimentally.

Here, we develop a microscopic model for long-time network relaxation that is controlled by cross-link dynamics. This \emph{cross-link governed dynamics} (CGD) model describes the structural relaxation that results from many independent unbinding and rebinding events. Using a combination of Monte Carlo simulations and an analytic approach, we demonstrate that this type of cross-link dynamics yields power-law network rheology arising from a broad spectrum of relaxation rates.  Our predictions are in excellent quantitative agreement with experiments performed on actin networks cross-linked with the transient linker protein $\alpha$-Actinin-$4$.

\begin{figure}[t]
\begin{center}
\includegraphics[width=.8\columnwidth]{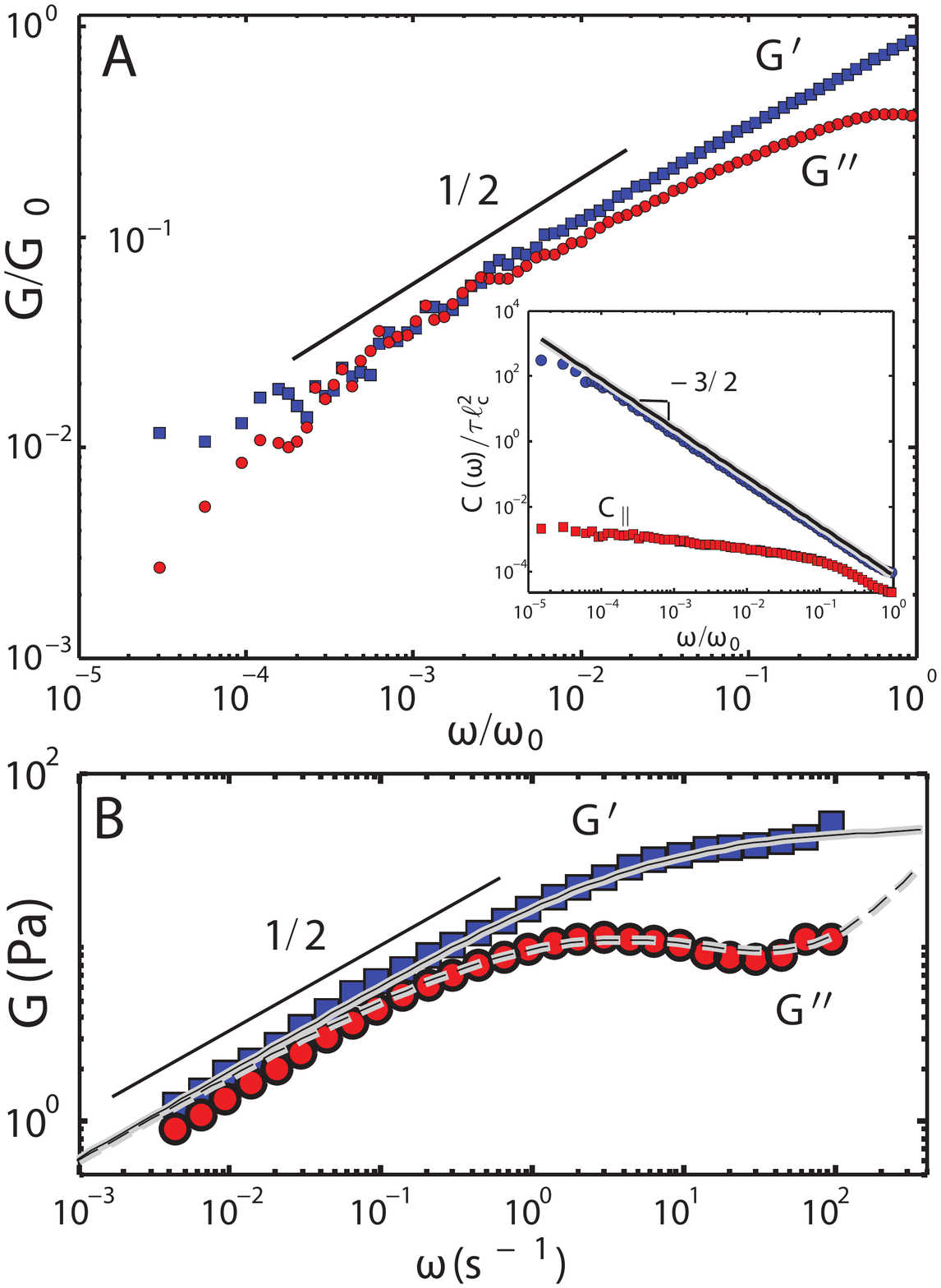}
\caption{ \label{fig:rexp} (Color online) A) The simulated rheology for frequencies below $\omega_0$. The shear modulus is normalized by the elastic plateau value  $G_0$ . The inset shows the total powerspectrum $C(\omega)$ (blue circles) of distance fluctuations, as well as the fraction $C_\parallel$ coming from effective stretch fluctuations originating in undulations on length scales shorter than the cross-linking distance. The distance fluctuations are determined over a length $16 \lc$ of a polymer with a persistence length $\lp=32 \lc$ and a total length $32 \lc$. The solid red line represents our analytical mean-field prediction.
B) Measured linear rheology of a 23.8 $\rm{\mu M}$ actin network cross-linked with 0.238 $\rm{\mu M}$ $\alpha$-Actinin-4. The low frequency behavior  is consistent with $G\sim (\imath \omega)^{1/2}$. The solid and dashed lines are global fits utilizing our mean-field CGD model for the low frequency regime together with the known high frequency response~\cite{gittes_dynamic_1998,morse_viscoelasticity_1998}.}
\end{center}\vspace{-0.2in}
\end{figure}

The CGD model can be qualitatively understood in simple physical terms. We assume each filament to be cross-linked to the network, with an average spacing $\lc$. Only filament bending modes between cross-links can relax (Fig.~\ref{fig:model}, lower inset), and the thermalization of these modes results in an entropic, spring-like response. To account for transient cross-linking, we assume that the linkers unbind at a rate $1/\toff$ (Fig.~\ref{fig:model},upper inset); this initiates the relaxation of long wavelength ($>\lc$) modes, giving rise to a softer spring constant and reduced macroscopic modulus. However, the relaxation of successively longer wavelength modes becomes slower, as an increasing number of unbinding events are needed for such a relaxation. This simple physical picture suggests a broad spectrum of relaxation times, as opposed to the single relaxation time of the Maxwell model. As outlined below both simulations and an analytic treatment of this model yield power-law behavior with $G\sim\omega^{1/2}$ below the characteristic frequency $\omega_0=2\pi/\toff$ (Figs.~\ref{fig:rexp} A and B).

We compare the basic predictions of this model to the rheology of a representative transiently cross-linked actin network. As a cross-linker, we use $\alpha $-Actinin-$4$ \cite{wachsstock_cross-linker_1994,miyata_strength_1996}, whose unbinding time $\toff$ is reported to be in the range $1 - 10 \rm s$, similar to that of other biological cross-linkers. These gels \cite{ward_dynamic_2008, bnorm} exhibit a low-frequency elastic shear modulus $G'$ with a pronounced decay over three decades in frequency, while the viscous modulus $G''$ exhibits a broad local maximum located near the characteristic frequency of cross-link unbinding~\cite{bnorm,ward_dynamic_2008,lieleg_transient_2008} (Fig.~\ref{fig:rexp}B). In the asymptotic low-frequency range, both moduli exhibit power-law rheology with an approximate exponent of $1/2$, in agreement with our predictions. Such behavior clearly indicates a more complex stress relaxation than captured by the Maxwell model, which is governed by a single relaxation time (Fig.~\ref{fig:model}). Taken together, the theoretical and experimental results demonstrate a distinct cross-link governed regime of network dynamics.

To develop a predictive microscopic model, we first consider a single polymer within the network, and then extend the description to the macroscopic level. On lengthscales longer than $\lc$, the motion of the polymer is constrained by its cross-linking to the surrounding network (see insets Fig. \ref{fig:model}). When a linker unbinds, a local constraint is released, allowing for the thermal relaxation of the freed segment. This thermal relaxation occurs within a time $\teq$, which is typically of order milliseconds~\cite{koenderink_high-frequency_2006,gittes_dynamic_1998,morse_viscoelasticity_1998}. We assume that this process is completed before the segment rebinds to the network at a new location; thus, $\teq \ll \ton$, where $\ton$ is the rebinding time of the linkers. Furthermore, assuming $\ton \ll \toff$, only a small fraction of cross-links will be unbound at any given time, and simultaneous unbinding of neighboring cross-links can be neglected. This suggests a coarse-grained description on length-scales longer than $\lc$, in which independent unbinding events occur at a rate $1/\toff$. Since the relaxation of wavelengths less than $\lc$ occurs at a much faster rate $1/\teq$, we use the worm-like chain model,
where the equilibrated short wavelength fluctuations manifest themselves as  an entropic stretch modulus $\muth \sim \kappa^2 / \lc^3 \kT $~\cite{mackintosh_elasticity_1995,morse_viscoelasticity_1998}. Here, $\kappa$ is the bending rigidity, $k_{\rm B}$ is the Boltzmann constant and $T$ is the temperature. In this description the coarse-grained energy is given by
\begin{equation}
\label{eq:Fcg} H_{CG}=\frac{1}{\lc}\sum_{n}\left[\frac{\kappa}{2}\left|\Delta
\bf{t}_n\right|^2+\frac{\muth}{2}\left(\left|\Delta {\bf
r}_n\right|-\lc\right)^2\right],
\end{equation}
where the sum extends over all cross-link positions ${\bf r}_n$, ${\bf t}_n$ is the unit tangent vector, and $\Delta$ represents a discrete difference.

Using $H_{CG}$, we study the dynamics arising from multiple linker unbinding and rebinding events, by performing 2D simulations of a single polymer. An initial chain conformation with periodic boundary conditions is randomly drawn from a Boltzmann distribution. Cross-link unbinding events are independent and result in the complete thermal equilibration of the two neighboring polymer segments. This is numerically implemented via a Metropolis Monte Carlo algorithm.

These simulations allow us to determine the equilibrium fluctuations of a single polymer, treating its surrounding network as a rigid medium. According to the fluctuation dissipation theorem (FDT), the linear mechanical response of the polymer is encoded in the fluctuations of the extension, $\delta\ell$, of the polymer. Interestingly, the simulations demonstrate that the power spectrum $C(\omega)=\langle |\delta \ell (\omega) |^2 \rangle$ depends on frequency as a fractional power-law, as shown in the inset of Fig.~\ref{fig:rexp}A, indicating a broad underlying distribution of relaxation times. The exponent is consistent with $-3/2$ over five decades in frequency. Although this exponent also arises in the Rouse model for flexible polymers due to the viscous dynamics of \emph{longitudinal} stretch modes~\cite{doi_theory_1999}, this is not the origin of the behavior found here. Our model does exhibit effective longitudinal stretch modes; however, their contribution $C_\parallel$ to the fluctuation spectrum is subdominant, as shown in the inset of Fig.~\ref{fig:rexp}A. This demonstrates that the polymer's response to an applied tension is dominated by the dynamics of long wavelength transverse modes.

The dynamical description of a single polymer can be extended to the network level by assuming that the network deforms affinely.
The macroscopic shear modulus $G^*$ is then related to the complex response function  $\chi$ of relative length extension  of a single polymer in response to a tensile force:
$G^*=\rho  / (15\chi)$,
where $\rho$ is the length of polymer per unit volume~\cite{gittes_dynamic_1998,morse_viscoelasticity_1998}. Ignoring end effects, the relative extension $\delta \ell/\ell$ of a polymer segment of length $\ell$ is conjugate to the uniform tension $f$, with $\delta \ell(\omega)/\ell=\chi(\omega) f(\omega)$. We use the FDT to calculate the imaginary part of the extensional response function \mbox{$\ell \chi ''(\omega)=\omega \langle \delta \ell^2 (\omega)\rangle /2 \kT$}. Using a Kramers-Kronig relation, we compute the full complex response function $\chi$, required to obtain the network shear modulus~\footnote{The Kramers-Kronig relation involves an integral over the whole frequency domain. Since we only simulate the low frequency part, we supplement this  with the expected plateau above $\omega_0$.}.
 Remarkably, below $\ooff$, the shear modulus depends on frequency as a power-law with an exponent of $1/2$ (Fig.~\ref{fig:rexp}A), consistent with experiments (Fig.~\ref{fig:rexp}B).

To obtain further insight into this behavior, we develop a continuum analytical treatment of this model. We calculate the polymer displacement due to the unbinding and subsequent rebinding of a linker to the $n$-th cross-link site along the chain backbone. To capture this equilibration step, we approximate the thermal distribution as Gaussian and centered around the mechanical equilibrium of the coarse-grained chain. This allows us to separate the equilibration step into a deterministic move to the minimum energy position together with a stochastic thermal contribution. The deterministic part of the displacement $\delta {\bf r}_n={\bf r}^{\rm (eq)}_n-{\bf r}^{\rm (i )}_n$ from the initial (i) position to the local equilibrium position (eq) is determined by
\begin{equation}
\label{eq:minimize}
{\bf 0}=\left. \frac{\partial { H_{\rm CG}}}{\partial {\bf r}_{n}}\right|_{{\bf r}_n={\bf r}^{\rm (eq)}_n}.
\end{equation}
This condition replaces the usual equation of motion balancing drag and conservative forces in the low-Reynolds number regime. In this continuum long-wavelength description, the leading order evolution equations are
\begin{eqnarray}
\label{eq:evolution1}
\toff \partial_t r_\parallel &=&\frac{\lc^2}{2} \partial_x^2  r_\parallel+\hat{x} \cdot {\boldsymbol{\xi}}_\perp\\
\label{eq:evolution2}
\toff\partial_t {\bf r} _\perp &=&\frac{\lc^2}{2} \partial_x^2 {\bf r}_\perp+{\boldsymbol{\xi}}_\perp,
\end{eqnarray}
where ${ \bf r}_\perp$ and $r_\parallel$ are the transverse and longitudinal deflections of the polymer with respect to its average direction $\hat{x}$.  The noise $\boldsymbol{ \xi}_\perp$ captures thermal effects, including local bucking due to thermally-induced compression~\cite{long_transient_paper}. It can be calculated from a quadratic expansion of $H_{CG}$ around its local mechanical equilibrium. For an inextensible polymer, the longitudinal component of this noise is subdominant and has been neglected.

\begin{figure}
\begin{center}
\includegraphics[width=0.92 \columnwidth]{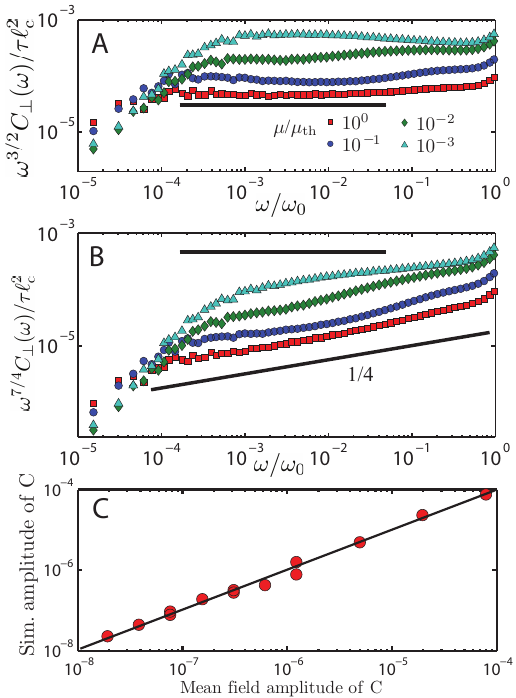}
\caption{\label{fig:scale} (Color online) The power spectrum $C_\perp(\omega)$ of longitudinal fluctuations originating from transverse undulations for length scales longer than $\lc$, multiplied with $\omega^{3/2}$ (A) and $\omega^{7/4}$ (B) for a range of polymer backbone compliances. C) The simulated amplitude of the power spectrum $C(\omega)/\tau \ell_c^2$ plotted against the 2D mean-field prediction for a range of polymer lengths and bending rigidities.}
\end{center}\vspace{-0.2in}
\end{figure}

Importantly, the noise $\boldsymbol{ \xi}_\perp$ depends nonlinearly on the local state of the polymer and couples Eqs.~(\ref{eq:evolution1},\ref{eq:evolution2}). To explore this coupling, we artificially reduce the stretch modulus $\mu$. In the limit $\mu\ll\muth$, the equations decouple and become exactly solvable, and the resulting transverse contribution to the fluctuation spectrum approaches $C_\perp\sim\omega^{-7/4}$. This can also be seen in our simulations with variable $\mu<\muth$ in Figs.~\ref{fig:scale} A and B. As $\mu$ is reduced below $\muth$, $C_\perp$ evolves toward $C_\perp\sim\omega^{-7/4}$, which can be seen by the flattening of the normalized spectrum in Fig.~\ref{fig:scale}B. In the limit $\mu\ll\muth$, the transverse bending dynamics are effectively those of a stiff filament fluctuating in a viscous solvent, for which the time-dependent fluctuations are $\langle|\delta\ell(t)|^2\rangle\sim t^{3/4}$\cite{granek_semiflexible_1997,gittes_dynamic_1998,morse_viscoelasticity_1998}. Only in this decoupled limit, can one understand the dynamics within the framework of an effective viscosity provided by the transient cross-links~\cite{long_transient_paper}.

The nonlinear nature of the effective noise precludes a full analytical solution of the model in the limit of an inextensible polymer. Instead, further insight is gained by approximating the amplitude of the effective noise term by its mean-field value calculated from the equilibrium fluctuations of the polymer. In this approximation, the stochastic contributions are uncorrelated in both time and space, resulting in the response function \cite{long_transient_paper}
\begin{eqnarray*}
\label{eq:alphamf}
\chi_{\rm MF}(\omega) &\approx& 0.0036 \frac{\kT \lc^3}{\pi\kappa^2} \int  \frac{\d q}{q^2-2\imath \omega \toff}.  \nonumber \\
\end{eqnarray*}
This response function captures the cross-link governed dynamics dominating on timescales $>\toff$. Further, we calculate the mean-field correlator, $C_{\rm MF}\sim\omega^{-3/2}$, in good agreement with the simulations presented in the inset of Fig.~\ref{fig:rexp}. As a further test, we perform simulations over a wide range of $\kappa$ and polymer lengths $L$; the predicted amplitudes are in excellent agreement with simulated amplitudes, as shown in Fig.~\ref{fig:scale}C. This further validates the assumptions made in our analytical approach.

To obtain a description of the mechanical behavior on all time-scales, we combine the response due to cross-link governed dynamics with drag-limited short-time dynamics~\cite{gittes_dynamic_1998,morse_viscoelasticity_1998}. Remarkably, the model can be fit to the experimental data---over the full range of frequencies---with three parameters: the plateau modulus $G_0=53$Pa, the equilibration rate $\teq=0.02$s and the unbinding time $\toff=3.2$s (see Fig.~\ref{fig:rexp}). This provides strong evidence that the low-frequency rheology of actin networks with the physiological linker $\alpha$-Actinin-4 is governed by the linker-controlled dynamics. Furthermore, the fitting procedure yields $\teq \ll \toff$; this, together with the quality of the fit, lends credence to the separation of timescales assumed in our model. Such a separation of timescales also implies that the fluid viscosity does not affect the dynamics or rheology in the linker-governed regime, consistent with observations in other experiments~\cite{lieleg_transient_2008}.

In addition to $\alpha$-Actinin-4, another known transient linker fascin has also been shown to yield network rheology consistent with $G^*\propto  (\imath \omega)^{1/2}$~\cite{lieleg_cross-linker_2007}. In fact, many physiological actin cross-linking proteins are dynamic and should exhibit a similar $\omega^{1/2}$ behavior. Indeed, this may enable the cell to regulate its response; on timescales short compared to the linker unbinding time, the network is effectively permanently connected---thereby providing mechanical resilience---while on longer timescales, dynamic linkers allow for complex network flow. This ability to flow and remodel is required for many vital cellular functions, ranging from motility to division. The extent to which transient cross-linking affects the mechanical properties of the cell is, however, still largely unknown. Interestingly, some rheological measurements on living cells have suggested a 1/2 power-law  behavior on time-scales ranging from several seconds to hours, consistent with the predictions of our model for a transient network~\cite{overby_novel_2005,desprat_creep_2005}. Further experimental work is needed to determine whether this regime is due to the transient nature of the cross-links.

\begin{acknowledgments}
This work was funded by FOM/NWO, the NIH (DK59588), the NSF (DMR-1006546), the Harvard MRSEC (DMR-0820484), and the DOE (FG02-97ER25308). We thank G. Barkema for useful discussions.
\end{acknowledgments}
\bibliography{TL}
\end{document}